# Potential Enabling Technologies for 7G Networks: Survey

Savo Glisic, Senior member IEEE, Department of Physics, Worcester Polytechnic Institute, Massachusetts
(sglisic@wpi.edu; savo.glisic@ieee.org)

*Abstract*- Every new generation of mobile networks ($n^{et}$'s) brings significant advances in two segments, enhancement of the network's parameters within the legacy technologies and introduction of new technologies enabling new paradigms in designing the $n^{et}$s. In the first class of enhancements the effort is to increase data rates, improve energy efficiency, enhance connectivity, reduce data transmission latency etc. In the second class of innovations for 6G and 7G, we anticipate focus on optimum integration of advanced ML (AI in general) and quantum (q-) computing (QC) with the continuous interest in the satellite $n^{et}$'s for optimal q-key distribution (QKD). By introducing q-technology 7G will be able to speed up computing processes in the $n^{et}$, enhance $n^{et}$-security as well as to enable distributed QC, which is a new paradigm in computer sciences.

Using advanced networks as a basic ingredient of intersystem integration, here we focus only on the second segment of anticipated innovations in networking and present a survey of the subset of potential technology enablers for the above concept with special emphasis on the interdependency of the solutions chosen in different segments of the network. In Section II, we present several anticipated 6G/7G network optimization examples resulting in a new paradigm of network optimization indicating a need for QC and QC based optimization algorithms. In Section III we surveyed work on quantum cryptography and QKD. Advanced solutions on energy and decision latency efficiency in ML are discussed in Section IV. In Section V we come back to network optimization and propose explicitly optimization framework for satellite network topology optimization to facilitate energy efficient QKD.

*NOTE on the writing style: In this paper we use specific notation where some characteristic terms, often repeated in the text, are replaced with corresponding acronyms representing the original term and its derivatives (conjugations). This approach (compressed language) enables more precise characterization of the system ($s^{yst}$) processes and operations and a specific term sound more like a $s^{yst}$ parameter that can be used more efficiently throughout the text. While this opens new options for the $s^{yst}$ presentation the writing occasionally sounds like an AI synthesized text. We hope the readers will easily get used to this style. In anticipation of what is coming in the field of ML and AI, this approach of integration of classical language and language of acronyms, might be further studied to increase the efficiency of Human-AI communication, maybe in the long run resulting in HAI language. Light acronymization used in this paper, only for illustration purposes, may be further intensified. The depth of acronymization would depend on specific application.*

*NOTATION (Acronyms)*

$\mathcal{A}^{coll}$-collective attacks
$\mathcal{A}^{nn}$-annihilation
$a^{lg}$- algorithm
$a\mathcal{N}$- artificial neuron
$a^{prox}$- approximations, approximately
bF-basis functions
$\mathcal{B}$- bosonic
$b^{io}$- biological
$b^{split}$- beam splitter
bSt-basis state
c-classical
ciM-circuit model
$c^{irc}$- circuits
cM -covariance matrix
$\mathcal{C}$-communications, communicate
$c^{an}$- canonical
$\mathcal{C}^{comm}$ -commutation, commutative, commuting, commute
$\mathcal{C}^{bas}$-computational basis
$\mathcal{C}^{ha}$- channel
$\mathcal{C}^{ng}$-change
$\mathcal{C}^{lass}$-classification
$\mathcal{C}^{ompl}$-complexity
$C, C^{omp}$ -computing, computation(aly), computer
$c^{od}$- coding,
$c\mathcal{S}$- configuration space
$c^{v}$-complex valued
cvF-complex-valued function
$cv\mathcal{V}$-complex-valued vectors
$\mathcal{C}^{re}$ - creation
cSt-coherent state
cF-cost function
Cry- cryptography
$d^{comp}$ -decompositions
$d^{cod}$- decoding
$d\mathcal{C}$-density code
$\mathcal{D}$ - density
$d^{ploy}$- deployment
$d^{iff}$- differential
$d\mathcal{G}$ - discrete group
$d^{str}$- distribution
dl-deep learning
$\mathcal{D}^{m}$ -m-dimensional
eD -eavesdropping
eF-eigenfunctions
$\mathcal{E}$ -entanglement, entangled, entangling
$\mathcal{E}^{xp}$-experience
$\mathcal{E}d$- entanglement distribution
eCry – encryption
$\varepsilon bp$- error backpropagation
$e^{drop}$- eavesdropper, eavesdropping
$e^{m}$-electromagnetic
$e^{cod}$- encoding, encoded
$e^{corr}$ -error correcting/correction
eS-eigenspectra
eSt-eigenstate
$\mathcal{E}st$-estimation, estimating, estimated
eV-eigenvelue
$e\mathcal{V}$ -eigenvectors
$e^{xp}$-experimental(y), experiment, experimented
exp- exponential
F-function
$f^{rate}$ -firing rate
fF- feedforward
$fs\mathcal{O}$ - free-space optical
$fo^{per}$- field operators
$\mathcal{FT}$- Fourier transform
$\mathcal{G}$ -Gaussian
$g\mathcal{C}$-generalized coordinates
gM-generalized momentum
Gl-gates library
G-gradient
$g\mathcal{I}$ graph isomorphisms
$\mathcal{G}i$-group-invariant
$gr\mathcal{S}$-graph state
$g^{nrlz}$- generalization
$\mathcal{H}$- Hamiltonian
$\mathcal{H}^{eis}$-Heisenberg
h-osc-harmonic oscillator
het-heterodyne
hom-homodyne
$h^{ogen}$- homogeneous
$h^{tgen}$- heterogeneous
$\mathcal{HS}$ Hilbert space
$\mathcal{J}$- information
$\mathcal{J}^{mpl}$- implementations, implementing, implement(ed)
$\mathcal{J}^{proc}$-information processing
lG-logic gate
$\mathcal{L}$-Lagrangian
$\mathcal{L}e$-learning
lt-long term
$\mathcal{M}$-measurements, measuring, measured
$m^{em}$- memory
$\mathcal{M}^{om}$-momentum
$m^{odu}$- modulation, modulated
$m\mathcal{S}$ momentum space
$m\mathcal{V}$-momentum vector
$\mathcal{N}$ - neuron
$n^{et}$- network(ing)
nN -neural networks
$\mathcal{N}^{lin}$ -nonlinearity
nSci -neurosciences
$\mathcal{O}$-optimization, optimal(y), optimize
$o^{per}$- operator, operations
$o^{pti}$- optical
p- privacy
$p^{erf}$- performance
$p^{last}$- plasticity
$p^{zed}$- parametrized, parametrization
$p^{rob}$-probabilities
$\mathcal{P}d$-probability distribution, density
$\mathcal{P}^{os}$-position
$\mathcal{P}^{th}$-path
$\mathcal{P}^{au}$- Pauli
$ph\mathcal{S}$-phase spaces
$p^{sh}$- phase shift
$\mathcal{P}h$-phenomenological, phenomena
$\mathcal{P}^{col}$-protocol
$\mathcal{P}^{oly}$-polynomial
$p\mathcal{S}$ position space
$p^{+}$- positive, positivity
$p\mathcal{V}$-position vector
$\mathcal{P}$- Perceptron
q-quantum
$\mathcal{Q}$-quadrature
qb-qubits
QC-quantum computing
QKD-quantum key distribution
qG-quantum gate
qCry-quantum cryptography
$q\mathcal{N}$ - quantum neuron
$q^{zed}$- quantized
$\mathcal{R}$-routing
$\mathcal{R}^{conc}$-reconciliation
r-repeater
$r^{eg}$- register
$r^{all}$- resource (channel) allocation
$\mathcal{R}^{epr}$-representation(al), represent(ing,ed)
$rv\mathcal{V}$ -real valued vector
$r^{ev}$- reversible
$ir^{ev}$- irreversible
rl-reinforcement learning
$r\mathcal{C}$-rate coding, rate code
$\mathcal{S}$-symplectic
s-secure, secret
$\mathcal{S}^{sec}$-security
$s^{tell}$- satellite(s)
$\mathcal{S}^{epa}$- separability, separable
sFi- spike firing
$s\mathcal{G}$ -symmetry group
$sh\mathcal{P}$- shortest path
$\mathcal{S}^{chrö}$ -Schrödinger
$\mathcal{S}^{ch}m$ - Schmidt measure
Si- swarm intelligence
sinC-single-carrier
sl -supervised learning
$s\mathcal{N}$-spiking neurons
$s^{ctr}$- spectrum
$sq^{z}$- squeezing, squeezed
$\mathcal{S}t$-stabilizer, stabilized
ST-spike-timing
$s^{rate}$ -spikes rate
subC-subcarrier, multicarrier
subCh-subchannel,
$s^{ymm}$- symmetrization
$\mathcal{S}y$- synaptic, synapse
$s^{yst}$- systems
$s^{ynth}$- synthesizing, synthesis
$\mathcal{S}yp$- synaptic plasticity

*$^{-}\mathcal{S}y$ -presynaptic*
*$^{+}\mathcal{S}y$ -postsynaptic*
*STDP-spike-timing dependent plasticity*
*st-short term*
*$\mathcal{T}$-tensoring, tensor*
*$\mathcal{T}^{map}$-transition maps*
*thF-threshold functions*
*$t^{rans}$- transformations, transform, transformed*
*$t^{pos}$ -transposition, transpose*
*$t^{port}$ -teleportation, teleport*
*$t^{miss}$ -transmission, transmitted, transmit*
*$t^{mit}$- transmittance*
*U- unitary*
*u-universal*
*ul -unsupervised learning*
*$\mathcal{US}$-universal set*
*$\mathcal{V}$-vector*
*W-Wigner*
*wF-wave function*
*w-wireless*
*w/q- wireless/quantu*

## 1. INTRODUCTION

Every new generation of mobile $n^{et}$ 's relies on more and more sophisticated technology and optimal choice of $n^{et}$ components becomes a rather complex ($\mathcal{C}^{ompl}$) optimization ($\mathcal{O}$) process. Besides the constant work on improvements of legacy technologies used in the previous generation there is always an effort to introduce new technologies enabling new paradigms in $n^{et}$ design. For 6G/7G, we believe these technologies will include, but not be limited to,

1. *q-computing ($C^{omp}$) to speed up execution of network optimization algorithms ($a^{lg}$'s) and enable use of q-$C^{omp}$ based cryptography (Cry)*
2. *spiking neural networks (nN) to significantly improve energy efficiency of overall AI engaged in the $n^{et}$ control.*
3. *experience aided learning to reduce decision latency*
4. *q-networking to enable distributed q- $C^{omp}$, a paradigm already used by computer science.*

## II. EVOLUTION of 7G OPTIMIZATION PROBLEMS

In this section we indicate how the network optimization complexity will increase in 6G/7G ecosystem leading to dependency on quantum computing, enabling significant computing speedups, and quantum optimization algorithms enhancing the computing efficiency. The problem will be further reiterated in Section V by discussing the optimization of large-scale networks like global networks.

By providing connectivity between the components of complex systems (network of networks or system of systems) communication network becomes a basis for intersystem integration. In 5G terminology these applications are known as *verticals*. As an example, in complex distributed industrial systems, like energy production, storage and distribution, it communicates data for control of energy sources, provides data for prediction of environment and user behaviour, energy consumption, price, anticipated consumption in the next time slot, energy storage capacity etc.

*Fig 1. Generic Model of Complex Intelligent Production System*

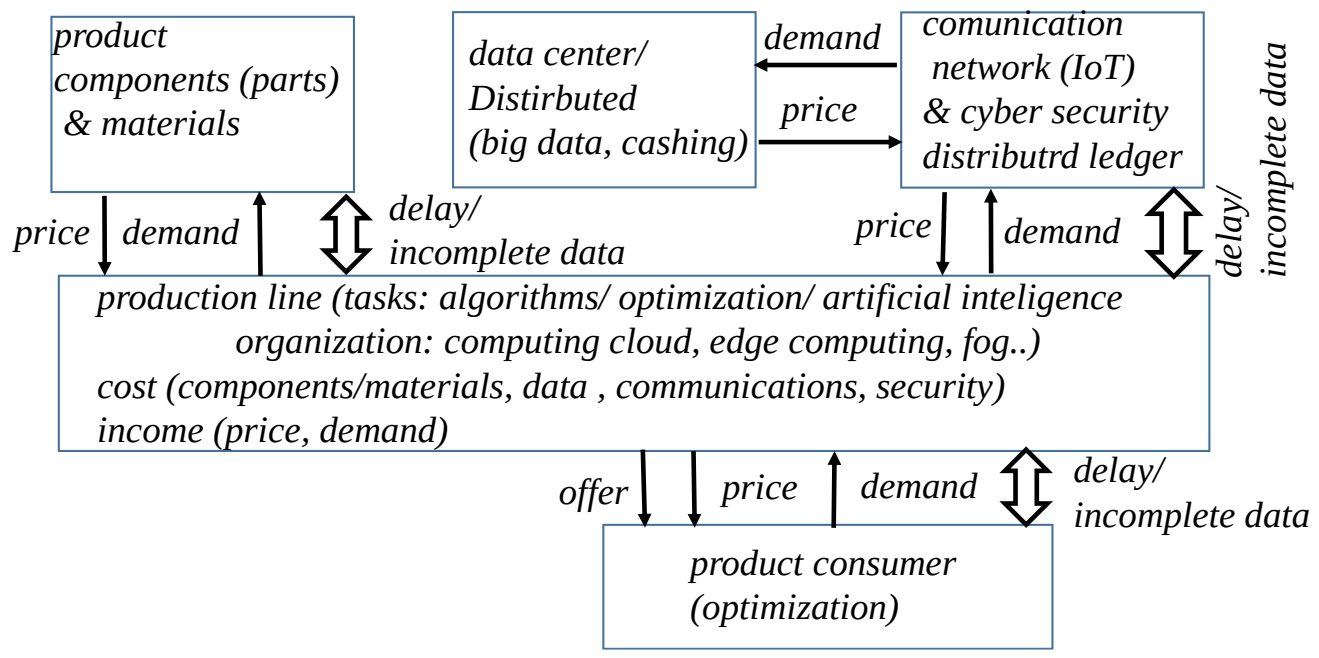

A generic model of a complex/intelligent production system, integrated by 7G, is shown in Fig.1. In the case of power grid, it can be instantiated to the model shown in Fig.1a. Similar models can be developed for autonomous driving, intelligent manufacturing, transportation network, online trading etc. A simplified model for joint multisystem optimization (next level of integration) like a *smart city* is illustrated in Fig.2. The major challenge here is to come up with a manageable objective function that can be efficiently optimized. Even at this stage it could be anticipated that due to the enormous number of variables and system dynamics, the optimization of such a complex objective function would be beyond the capacity of classic computers and would require capacities promised by quantum computers.

*Fig1.a EXAMPLE: Power Grid*

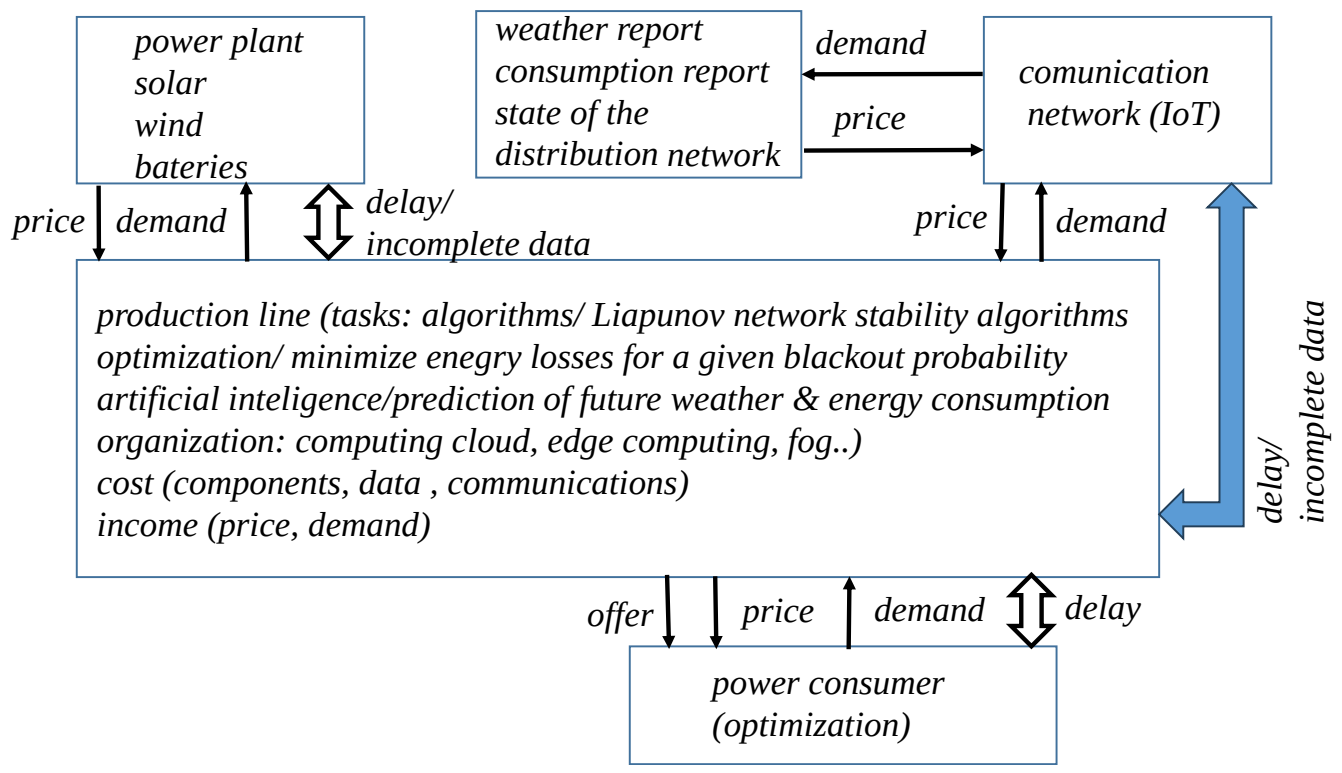

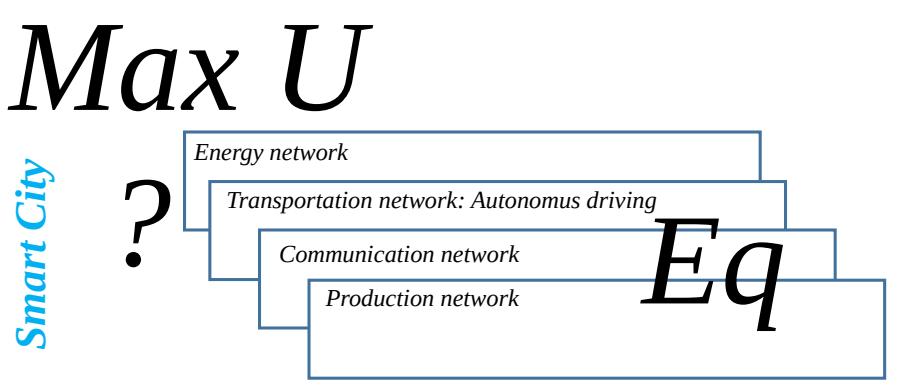

*Fig.2 Joint Optimization of multiple system model (Smart City)*

For a simple model where the objective functions, expressed in terms of revenue $r(i,\tau)$ and cost $c(i, \tau)$, depends only on information incompleteness $i$ and delay $\tau$ in the delivery of the information across the network, we can define different optimization problems.

For high precision in parameter estimation (power consumption, weather forecasting, storage capacity etc), the revenue would be high which requires low $i$. For this, sensor networks should be densely populated which translates into the high cost and amount of traffic generated would be higher demanding more sophisticated communication network. Similarly, for the same reason, to have fresh data, $\tau$ should be low, requiring high-rate data networks which are more expensive. This model can be further instantiated as shown in Fig.3.

At this model we have grouped all sources of energy from the same family of plants into one common pull of sources. In a real network these sources may be distributed over a large area requiring complex communication and power distribution networks requiring more detailed modeling.

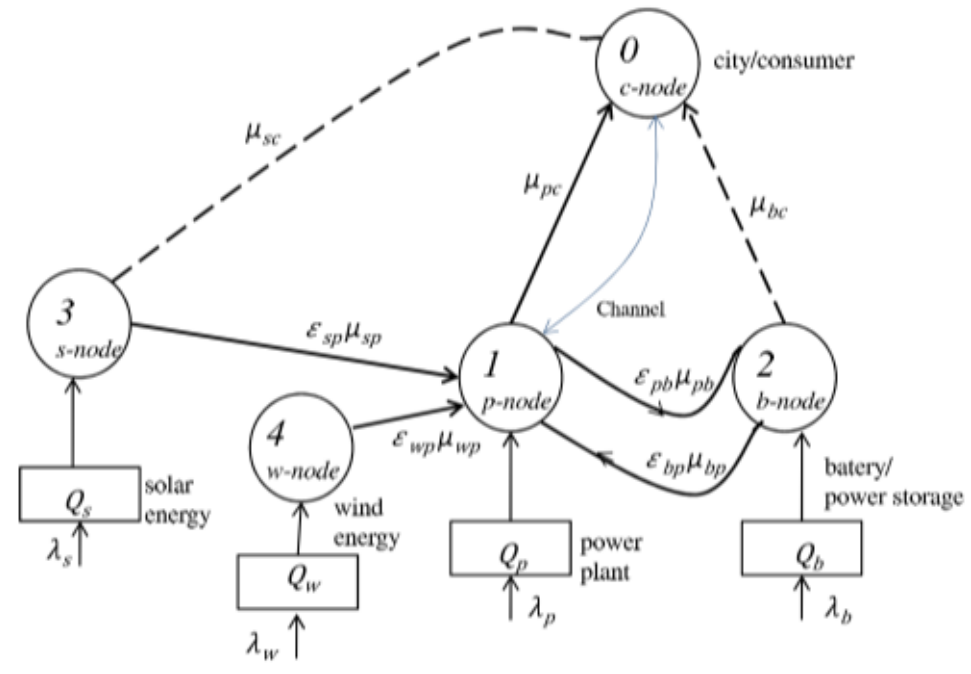

*Fig.3 Power grid model*

For a given prediction of energy production in nodes *s* (*solar*) and *w (wind)*, estimated storage capacities/electrical vehicles charging station (*b*) and estimated consumption (*c*), we can plan amount of energy production in the power plant (*node p*) to minimize over-production which are losses and prevent under-production resulting in blackout. Further elaboration of different models depends on existing network topologies, specific production processes and business models (choice of the optimization problem which is often in discretion of the process owners). For this reason, we leave to the reader to further elaborate these models and adopt the best approach for their specific problems.

We will revisit the problem of network optimization in the context of global network design in Section V.

Since communication networks and parameter estimation based on ML (AI) are key component in optimization of the system, in Section IV we provide to the researchers, network designers and planners of the 7G standards a brief survey of the candidate technologies enabling the above solutions with emphasis on latency and power consumption reduction.

## III. ROLE OF QUANTUM COMPUTING

When moving from 6G to 7G solutions we will also witness step by step transition from nano to quantum technology.

The advanced network solutions will be looking into the application of q-technology at least for two reasons, to use q-based cryptography and build up q-network for quantum key distribution and enable distributed q-computing by interconnecting near future quantum computers. At the same time, quantum computers, due to their computing speedups (up to factor $10^8$), will enable optimization of the problems with much more complicated objective functions, as indicated in the previous section, in dynamic systems where updates of the optimum solutions must be calculated more often. Most of these optimization problems are combinatorial in nature (NP hard) and Quantum Approximative Optimization Algorithms (QAOA) offer significant enhancement in computing efficiency [1]. Quantum search algorithms, like Grover's algorithm [1], also provide significant speed up in finding the optimum of an objective function. For a data set of size N, classical exhaustive search would require in average $\sim$N trials to find the optimum point while Grover's algorithm would require $\sim\sqrt{N}$ trials. Even in the simple case of N=$10^6$ instead of searching million points in the data set, Grover's algorithm would need only $10^3$ trials.

A q-$s^{yst}$ can be modeled using discrete (DV) or continuous variable (CV). While discrete variable approach [1] provides a better platform for systematic introduction to QC, continuous variable solutions are more feasible options for practical implementation in 7G.

*A. Quantum Cryptography and QKD*

Classical cryptography ($c^{rypt}$) schemes ($s^{che}$) have been compromised by the practical results on quantum ($q$) computers in recent years. Nowadays these $s^{che}$ 's can be compromised by using the Shor's methodology. This paper provides a brief survey of the work on so called post- $q$ $c^{rypt}$ (PQC) $s^{che}$ 's, based on different principles, minimizing the threats coming from advances of $q$-computers. Even so, post-$q$-$s^{che}$ 's do not completely solve the problem ($p^{rblm}$) but rather represent ($r^{prs}$) a temporary solution. On the other hand, $q$-$c^{rypt}$ (QC) and $q$- key distribution ($d^{istr}$) (QKD), discussed in this paper, offer the ultimate solution: by relying on entanglement ($\mathcal{E}^{gle}$) between $q$- states ($\mathcal{S}^{tat}$'s). At least in the beginning, a competition is anticipated between the two approaches to security (S) $s^{che}$ 's, so the paper provides a brief survey of both QC and PQC algorithms ($\mathcal{A}^{lgrt}$'s), enabling full understanding of pros and cons when choosing implementation ($\mathcal{J}^{mpl}$) options in future networks ($n^{et}$'s). For more details see [2].

*Post-q Cryptography*

The legacy $c^{rypt}$ -$s^{che}$ 's depend on integer ($i^{nteg}$) factorization or discrete ($D$) logarithm $p^{rblm}$, which have been compromised by using the Shor's $\mathcal{A}^{lgrt}$ [1]. As a result, active research has been conducted to find solutions based ($b^{se}$'$d$) on different principles, which are minimizing the threats coming from advances in $q$-computers ($c^{ompt}$) referred to as *post- q- $\mathcal{S}$- $s^{che}$ 's.*

Within this umbrella a number ($n^{umb}$) of different classes of solutions has been published including: *Mv- $c^{rypt}$:* using polynomials ($p^{oly}$'s) with multiple variables ($v^{ari}$'s) ($Mv$) over a finite field ($\mathcal{F}$) F ($d^{fin}$'ed in ground or expansion $\mathcal{F}$). Solving such $p^{rblm}$ 's is either NP-hard or NP-complete. Therefore, they are strong contenders of post- $q$- $c^{rypt}$. *Mv- $c^{rypt}$* has one very important advantage since it uses very short signature [3], which can serve the purpose of authentication in small devices ($d^{vic}$'s). An example in this class is *Rainbow*, a signature $s^{che}$ [4]. Here we briefly list work on three classes of these $\mathcal{A}^{lgrt}$ 's:

1. *Lattice ($\mathcal{L}^{att}$) -$b^{se}$ 'd -$c^{rypt}$*: The first version of the system ($s^{yst}$), called NTRU (**N**-th $d^{gre}$ **T**runcated $p^{oly}$ - **R**ing ($\mathcal{R}$) **U**nits) , was developed in [5]. The latest variant of NTRU or NTRU Prime [6], is stronger than the original NTRU by creating stronger algebraic structure. Additional versions in this class are *Learning with Errors ($\in$) over $\mathcal{R}$'s* ($\mathcal{R}$-LWE) [7] and *BLISS (Bimodal $\mathcal{L}^{att}$ Signature $s^{che}$* (BLISS) [8]
2. *Hash ($\mathcal{H}^{sh}$) $b^{se}$ 'd -$c^{rypt}$*: Examples in this class are *Merkle* [9] and *Lamport Signature* [10].
3. *Code $b^{se}$ 'd -$c^{rypt}$* includes $s^{che}$ 's like McEllice [11], and Niederreiter crypto- $s^{yst}$.

While the post- $q$-$c^{rypt}$ offers near term ($t^{rm}$) solutions (most probably a prevailing implementation ($\mathcal{J}^{mpl}$) in 6G $n^{et}$ 's) an extensive work in the research community is focused on $q$-$c^{rypt}$ that should $r^{prs}$ an ultimate solution (foreseen for $\mathcal{J}^{mpl}$ in 7G- $n^{et}$ 's).

*q- Cryptography*

For the basics in $q$- $c^{ompt}$ see Appendix of this paper and [1]. Here we consider ($c^{nsdr}$) both, $D$ and continuous ($C^{ont}$)- $v^{ari}$ -$s^{yst}$ 's [1,2,12 ]. First, we summarize some preliminary aspects used in the $m^{dl}$'ing and analysis of both types of the $s^{yst}$ 's.

*Prepare ($p^{rep}$) and measure QKD $\mathcal{P}$* consists of two parts: $q$- communication ($c^{omm}$) ($qC$) and $c$- post- $p^{rces}$ 'ing ($cP$) [2]. During $qC$ the $t^{rns}$ 'er (node A) encodes ($\mathcal{E}^{cod}$) a $r^{ndm}$ $c$-signal ($s^{gnl}$) $\alpha$ into non- $o^{rth}$ -$q$- $\mathcal{S}^{tat}$ 's which are then sent over a $q$-$c^{han}$ controlled by the eavesdropper (node E), who tries to steal the $\mathcal{E}^{cod}$ 'ed information ($\mathcal{J}$). The $l^{ine}$ 'ity of $q$- mechanics ($m^{chn}$) do not allow to perform ($p^{rfm}$) ideal cloning ($c^{lon}$'ing) [ 13 ], so that E can only get partial $\mathcal{J}$ while disturbing the $q$-$s^{gnl}$ 's. At the $o^{tp}$ of the $c^{omm}$ -$c^{han}$ ($cC$), node B- $\mathcal{M}$'s the received $s^{gnl}$ 's and gets a $r^{ndm}$ $c$- $s^{gnl}$ -$\beta$. After several uses of the $c^{han}$, node A and node B share raw data $r^{prs}$*'ed* by two correlated $v^{ari}$ 's $\alpha$ and $\beta$.

The two nodes use portion of the data to evaluate the $p^{mtr}$ 's of the $c^{han}$, its $t^{miss}$ and noise ($\mathcal{N}$), needed to estimate how much of post- $p^{rces}$ 'ing is needed to distill a $pK$ from the rest of the data. Depending on this $\mathcal{J}$, they do $\in$ correction ($c^{rct}$) (eC), enabling them to detect ($d^{tct}$) and remove $\in$'s, followed by a $p^{rces}$ of privacy ($p^{riv}$) amplification ($a^{mpl}$) designed to decrease $\mathcal{J}$ stolen by E's. The result is the $s$- key ($sK$). We have direct reconciliation (DR) or reverse reconciliation (RR), depending on which $v^{ari}$ is inferred.

In DR, node B does post- $p^{rces}$ 'ing of its outcomes to infer A's $\mathcal{E}^{cod}$ 'ings. This $p^{rces}$ is usually assisted by forward ($f^{orw}$) $c^{rect}$ 'ing code (CC) from node A to node B. By contrast, in RR, it is node A who post- $p^{rces}$ 'es her $\mathcal{E}^{cod}$ 'ing $v^{ari}$ to infer B's outcomes. This can be enhanced by a $b^{kwr}$ -CC from node B to node A. One may more $G$ 'ly use two-way $p^{rces}$ where the extraction of the $K$ is assisted by $f^{orw}$ and feedback CCs, which may be even interleaved with the various $c^{omm}$ rounds of the $\mathcal{P}$. Asymptotic ($a^{symp}$) $\mathcal{S}$ of such $s^{che}$ 's is studied in [14].

*Composable ($\mathcal{C}^{omp}$) $\mathcal{S}$ of QKD:* $c^{rypt}$ 'ic tasks are often incorporated into larger $\mathcal{P}$'s. With QKD secure ($s$) $c^{omm}$ can be constructed by using $K$- $d^{istr}$ together with the one-time pad $\mathcal{P}$. If two $\mathcal{P}$'s are $s$ in accordance with a $\mathcal{C}^{omp}$- $\mathcal{S}$- $d^{fin}$ 'ition, then their amalgam is $\mathcal{S}$ without the need to give a separate $\mathcal{S}$ proof ($p^{rf}$) for the composite $\mathcal{P}$ [15].

*Discrete $v^{ari}$ -$\mathcal{P}$'s*

DV $\mathcal{P}$'s are the simplest $f^{rm}$ of QKD like BB84 (**B**ennett and **B**rassard, 1984) $\mathcal{P}$. For the beginnings of $q$- $c^{rypt}$ see [ 1 ,16 ]). Some basic background needed in this section is summarized in the Appendix.

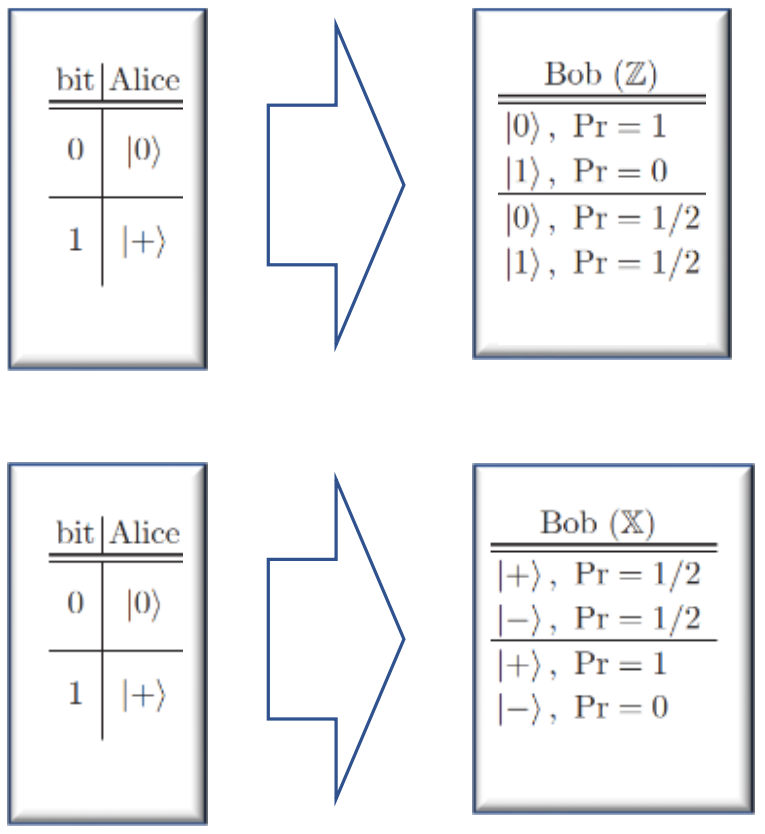

*Fig. 4. B's outcomes and their probabilities (Pr) depending on A's $\mathcal{E}^{cod}$ 'ing $\mathcal{S}^{tat}$ and B's measurement ($\mathcal{M}$) $b^{se}$*

*B92 $\mathcal{P}$*: C. **B**ennett, proposed in 19**92**, a simple $\mathcal{P}$ of QKD, the "B92" [2 ]. By employing two $\mathcal{S}^{tat}$ 's it distributes a $sK$ between the nodes. This is the minimum needed for transmitting ($t^{rns}$*'ing*) one bit of a $c^{rypt}$ 'ic $K$ ($cK$). Here, node A $i^{nit}$ 's a qubit ($qb$) in one of two $q$- $\mathcal{S}^{tat}$ 's, $|\psi_0\rangle$ and $|\psi_1\rangle$, by associating to them 0 and 1, $r^{spct}$ 'ively. The $\mathcal{S}^{tat}$ is $t^{rns}$ 'ed to nide B, who $\mathcal{M}$'s it in a proper $b^{se}$, to recover the bit sent by node A. If the $\mathcal{S}^{tat}$ 's $|\psi_0\rangle$, $|\psi_1\rangle$ were orthogonal ($o^{rth}$), node B would be always able to $c^{rect}$ 'ly recover the bit. For example, if $|\psi_0\rangle = |0\rangle$ and $|\psi_1\rangle = |1\rangle$, node B can $\mathcal{M}$ the received $\mathcal{S}^{tat}$ 's in the $\mathbb{Z}$- $b^{se}$ ($|0\rangle, |1\rangle$) and $d^{tct}$ the information ($\mathcal{J}$) with 100% $p^{rob}$. On the other hand, B's ability to recover the $\mathcal{J}$ without any error also means that E can do it as well. E will $\mathcal{M}$ the $\mathcal{S}^{tat}$ 's between node A and node B, $c^{rect}$ 'ly recover the $\mathcal{J}$, $i^{nit}$ new $\mathcal{S}^{tat}$ 's same as the $\mathcal{M}$'ed ones, and $t^{rns}$ them to node B, who cannot notice any difference from the $\mathcal{S}^{tat}$ 's sent by node A. The $o^{rth}$ -$\mathcal{S}^{tat}$ 's behave like $c$- ones, which can be $c^{rect}$ 'ly $\mathcal{M}$'ed, copied and cloned. The $o^{rth}$ -$\mathcal{S}^{tat}$ 's are eigen-$\mathcal{S}^{tat}$ 's of some common observable, and $\mathcal{M}$'s made using that observable would not be subjected to any uncertainty ($u^{crt}$). The no- $c^{lon}$ theorem ($t^{eor}$) [17 ] does not apply here. By contrast, $\mathcal{M}$'s will be $b^{nd}$ 'ed by inherent $u^{crt}$ 's if node A -$\mathcal{E}^{cod}$ the $\mathcal{J}$ in two non- $o^{rth}$ -$\mathcal{S}^{tat}$ 's, like: $|\psi_0\rangle = |0\rangle$, $|\psi_1\rangle = |+\rangle$, $\langle\psi_0|\psi_1\rangle = s \neq 0$, where $p^{mtr}$ -s , is $\mathcal{O}^{ptmz}$ 'ed to provide the

best performance ($p^{erf}$) of the $\mathcal{P}$. For the above $\mathcal{S}^{tat}$ 's, this $p^{mtr}$ is selected as $1/\sqrt{2}$. Given the complementary character of the observables involved in differentiation between these $\mathcal{S}^{tat}$ 's, neither B nor E can $\mathcal{M}$ or copy the $\mathcal{S}^{tat}$ 's sent by A with $p^{rob}$ equal to 1. The $\mathcal{S}$ of the B92 $\mathcal{P}$ is $b^{se}$ 'd on the fact that, while A and B can easily mitigate this $p^{rblm}$ (see below) and distil a common bit from the data, E cannot.

Recall that the $\mathcal{S}^{tat}$ $|0\rangle(|+\rangle)$ is an eigen- $\mathcal{S}^{tat}$ of $\mathbb{Z}$ ($\mathbb{X}$) and that $|\pm\rangle = (|0\rangle \pm |1\rangle)/\sqrt{2}$, see [1] or Appendix. Now, if node A $p^{rep}$ 's ($i^{nit}$) the $\mathcal{S}^{tat}$ -$|\boldsymbol{\psi_0}\rangle$ and node B- $\mathcal{M}$'s ($d^{tct}$ 's) it with $\mathbb{Z}$, he will see $|\mathbf{0}\rangle$ with $p^{rob}$ 100% while when he $\mathcal{M}$'s it with $\mathbb{X}$ ($|+\rangle, |-\rangle$), he will see either $|+\rangle$ or $|-\rangle$ with $p^{rob}$ 50%. Here node B will never see, $|\mathbf{1}\rangle$. On the other hand, if A $i^{nit}$ 's the other $\mathcal{S}^{tat}$ of B92, $|\boldsymbol{\psi_1}\rangle$, node B will still $d^{tct}$ in the same $b^{se}$ 's as before but this time, node B cannot see the $\mathcal{S}^{tat}$ $|-\rangle$ as a result. See Fig. 4 for a list of node B's outcomes and their $p^{rob}$*'s* depending on node A's -$\mathcal{E}^{cod}$ 'ing $\mathcal{S}^{tat}$ and B's $s^{lct}$ 'ed $\mathcal{M}$ $b^{se}$. From the figure, for the conditional $p^{rob}$ $p(A|B)$ of estimating A's -$\mathcal{E}^{cod}$ 'ing $A$ given B's outcome $B$, we may write $Pr(|+\rangle||1\rangle) = Pr(|0\rangle||-\rangle) = 1$ i.e. node B assumes that when he sees $|1\rangle$, node A must have $i^{nit}$ 'd the $\mathcal{S}^{tat}$ $|+\rangle$, and when he $d^{tct}$ 's $|-\rangle$, node A must have $i^{nit}$ 'd the $\mathcal{S}^{tat}$ $|0\rangle$. If any other $\mathcal{S}^{tat}$ is $d^{tct}$ 'ed, node B is unsure of what node A's has $i^{nit}$ 'd and the users remove these outcomes from their records.

So, by "reversed decoding", and his joint effort with A, B can decode the $\mathcal{J}$ sent by A. For more details see [2,18].

BB84 $p^{erf}$ 's better than the B92 $\mathcal{P}$. The use of only two linearly ($l^{ine}$*'ly*) independent ($i^{dep}$) $\mathcal{S}^{tat}$ 's enables $E$ to do an "unambiguous $\mathcal{S}^{tat}$ discrimination" USD $\mathcal{M}$ on the *q-* $\mathcal{S}^{tat}$ 's -$i^{nit}$ 'd by A. Even so, it was presented here for its simplicity.

Continuous research has confirmed that using CV instead DV q-information carriers, offers better performance ($p^{erf}$) of q-information ($\mathcal{J}$) processing ($\mathcal{J}^{proc}$) by using Gaussian ($\mathcal{G}$-) models for q- states, q- operations ($o^{per}$), and q-measurements ($\mathcal{M}$). Using these models makes both theory and experimentation ($e^{xp}$) more efficient. For theoretical work simple analytical tools are available, enabling closed form expressions for many $p^{erf}$ measures and on the $e^{xp}$-side $\mathcal{G}$-processes are easier to reproduce in the laboratory. $\mathcal{G}$-q-$\mathcal{J}^{proc}$ is used in q-communication ($\mathcal{C}$), q-*Cry*, QC, q-teleportation ($t^{port}$), and q-state and $\mathcal{C}^{ha}$ discrimination, all of which are of interest for 7G networks.

An example of DV q-$\mathcal{J}$ is the *qb* [1], a q- $s^{yst}$ with two distinct states like two lowest energy states of semiconductor q-dots. An example of CV q-$\mathcal{J}$ is the $q^{zed}$ harmonic oscillator (*h-osc*), modeled by CV's such as *pos*ition ($\mathcal{P}^{os}$) and *momentum* ($\mathcal{M}^{om}$). Examples of CV q-$s^{yst}$ 's are $q^{zed}$ modes of bosonic ($\mathcal{B}$) $s^{yst}$ 's like different degrees of freedom of the electromagnetic ($e^{m}$) field. For an example of creating entangled Bell states see Fig.5.

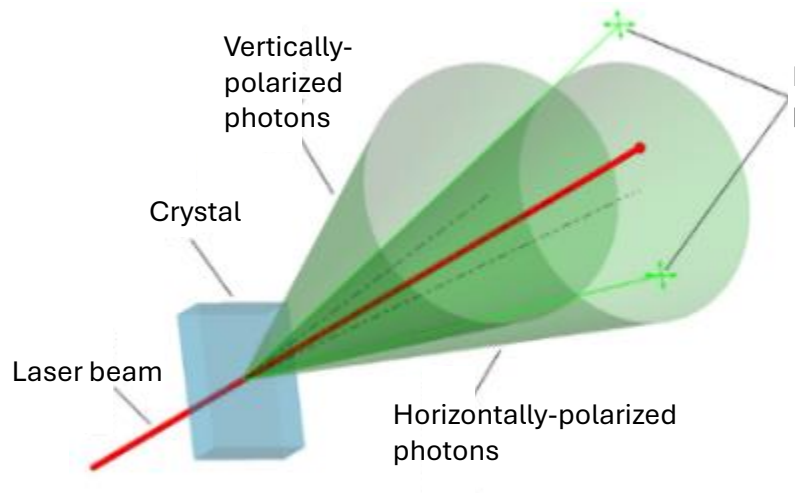

*Fig.5* An illustration of creating entangled Bell states *https://en.wikipedia.org/wiki/Bell_state*

## IV ENERGY AND DECISION LATENCY EFFICIENCY of ML in 7G

7G network designers should be interested in spiking neural networks for at least two reasons. First using spikes instead of continuous presence of signals enables several orders of magnitude better energy efficiency. Second, deep understanding of the neurological processes enables us better insights in the operation of the human brain which we expect to help us further develop better modelling and design of the $a^{lg}$ 's for control of artificial neural networks in 7G- n$^{et}$'s.

### *Spiking neuron*

Synaptic (*$\mathcal{S}$y)* plasticity (*$\mathcal{S}$yp)* (Fig.6) enables $b^{io}$-learning and memory. Here we survey works on modeling short-term (*st*) and long-term (*lt*) *$\mathcal{S}$yp*, with focus on *spike-timing (ST) dependent plasticity* ($p^{last}$) *(STDP)*. This provides a framework for analyzing various models of $p^{last}$.

We discuss *$\mathcal{S}$y* update rules for *st* or *lt*- $p^{last}$ that depend on *ST,* membrane potential, and the value of the 𝒮𝓎 weight. We further briefly review the literature discussing their relations to *sl* and reward-based rules (*rl*).

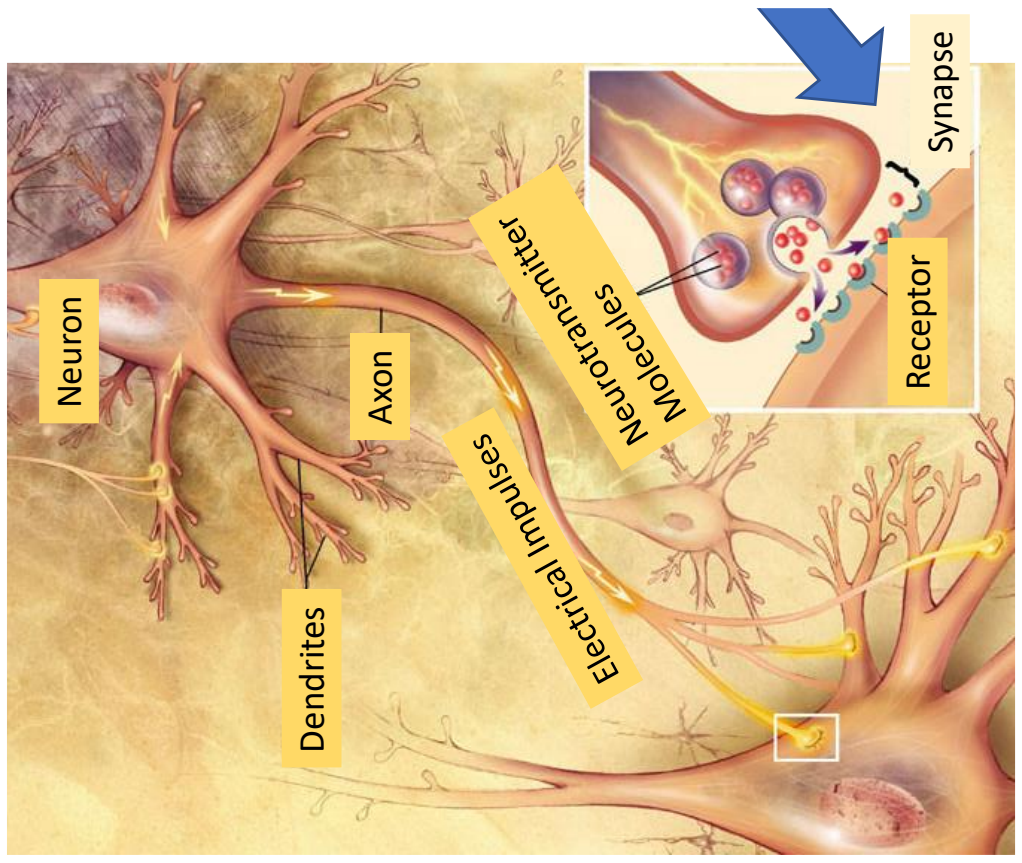

*Fig. 6. Artistic interpretation of the major elements in chemical synaptic transmission. In neuroscience, 𝒮𝓎p is the ability of synapses to strengthen or weaken over time, in response to increases or decreases in their activity [19]*

𝒮𝓎-$\mathcal{C}^{ng}$'s are involved in ℒℯ memory, and cortical $p^{last}$. In $e^{xp}$ -initializations, 𝒮𝓎-$\mathcal{C}^{ng}$'s can be inserted by an impact defined by presynaptic ($^{-}$𝒮𝓎) firing rates, postsynaptic ($^{+}$𝒮𝓎) membrane potential ($p^{ot}$), calcium entry, or *ST* [19].

Although biophysical models are important to describe the biological ($b^{io}$) mechanisms having impact on 𝒮𝓎p, phenomenological (𝒫𝒽) models representing the 𝒮𝓎 $\mathcal{C}^{ng}$ 's without reference to such mechanism are more tractable and simpler to implement. In the literature, a 𝒮𝓎 from a $^{-}$𝒮𝓎-𝒩 to a $^{+}$𝒮𝓎 -𝒩 are considered. The intensity of a connection from/to is described by a weight that depends on the magnitude of the $^{+}$𝒮𝓎 response, ℳ as the height of the $^{+}$𝒮𝓎 -$p^{ot}$ or the slope of the $^{+}$𝒮𝓎 current. Directions and magnitudes of $^{+}$𝒮𝓎-$\mathcal{C}^{ng}$'s can be stated as '$^{+}$𝒮𝓎 *update rules*' or '*ℒℯ- rules*'. Here we limit our interest to rules accounting for the results of ℳ in which 𝒮𝓎p was observed as a result of $^{-}$𝒮𝓎 and $^{+}$𝒮𝓎 spikes (for more details, see [20]). For specification of 𝒮𝓎p rules, we use the time needed to *induce* such $\mathcal{C}^{ng}$ and duration of *persistence* of the $\mathcal{C}^{ng}$. For *st* and *lt* plasticity, $\mathcal{C}^{ng}$ 's can be inserted in less than 1*s*.

In *st*- $p^{last}$, a series of eight $^{-}$𝒮𝓎 spikes at 20Hz evokes smaller (depression) or larger (facilitation) responses in the $^{+}$𝒮𝓎 cell. For *st*- $p^{last}$ this $\mathcal{C}^{ng}$ persists for less than a few 100 ms: the size of the $^{+}$𝒮𝓎 response recovers within less than a second. Different from *st*- $p^{last}$, *lt*-potentiation and depression (*lt*-P and *lt*-D) refer to persistent $\mathcal{C}^{ng}$ 's of 𝒮𝓎-responses. The time needed for induction can still be relatively brief. In *STDP*, a $\mathcal{C}^{ng}$ of the 𝒮𝓎 can be inserted by 60 pairs of $^{-}$𝒮𝓎 and $^{+}$𝒮𝓎 spikes with a repetition frequency of 20 Hz, completing the stimulation within 3s. On the other hand, this $\mathcal{C}^{ng}$ can last for more than one hour. The final stabilization of a potentiated 𝒮𝓎 follows only thereafter, referred to as the late phase of *lt-P*. The 𝒩's in the brain must remain within a sustainable activity regime, despite the $\mathcal{C}^{ng}$ 's inserted by *lt-P* and *lt-D*. This is achieved by homeostatic $p^{last}$, an up- or down-regulation of all *𝒮𝓎's* converging onto the same $^{+}$𝒮𝓎 𝒩 which occurs in a time segment between few minutes- few hours long [19].

The 𝒫𝒽 -models discussed in this section can be classified as *ul* rules. ℒℯ consists of adjusting the 𝒮𝓎 to the statistics of the activity of $^{-}$𝒮𝓎 and $^{+}$𝒮𝓎 𝒩's which is different from reward-based learning (*rl*). The relevance of signaling chains for models of 𝒮𝓎p as well as the importance of the $^{+}$𝒮𝓎 voltage, is also explored in the literature.

*A Spiking 𝒩- $n^{et}$ 's (SNN)*

SNNs are considered as the 3rd generation of NN's. Based on $C^{omp}$ paradigms in the brain and recent advances in neurosciences (*nSci*), they leverage an accurate modeling of 𝒮𝓎 interactions between *𝒩's*, considering the time of spike firing (*sFi*). The original work proposed a NN model based on simplified "binary" 𝒩's, where a single 𝒩- $\mathcal{J}^{mpl}$ a simple thresholding *F* (*thF*): a state of a 𝒩 is either "active" or "not active", and at each 𝒩- $C^{omp}$ step, this state depends on the weighted sum of the states of all the input 𝒩's that connect to the given 𝒩. Links between 𝒩's are directed (from neuron $\mathcal{N}_i$ to neuron $\mathcal{N}_j$) and have weight ($w_{ij}$). If the weighted sum of the states of all the 𝒩's $\mathcal{N}_i$ connected to a 𝒩 $\mathcal{N}_j$ is higher than the preset threshold of $\mathcal{N}_j$, the state of $\mathcal{N}_j$ is set to active, otherwise it is not (see Fig. 7).

Even a $n^{et}$ of such simple elements can $\mathcal{J}^{mpl}$ a range of mathematical $F$ ‘s. Different $\mathcal{L}e$ rules, both for teaching a $n^{et}$ to perform some assignments (*sl*), and for $\mathcal{L}e$ specific features “on its own” (*ul*) have been proposed. Gradient (*G-*) descent $a^{lg}$‘s (e.g. error backpropagation-$\varepsilon bp$ [1]) that guide the NN behavior to some target objective $F$, are an example of *sl* -$a^{lg}$ ‘s. Many works on local *ul* in *NN*’s originate from the work by Hebb. Such *ul* rules are known as Hebbian rules (e.g. in Hopfield’s $n^{et}$). Artificial NN’s are used, as engineering tools, in many domains

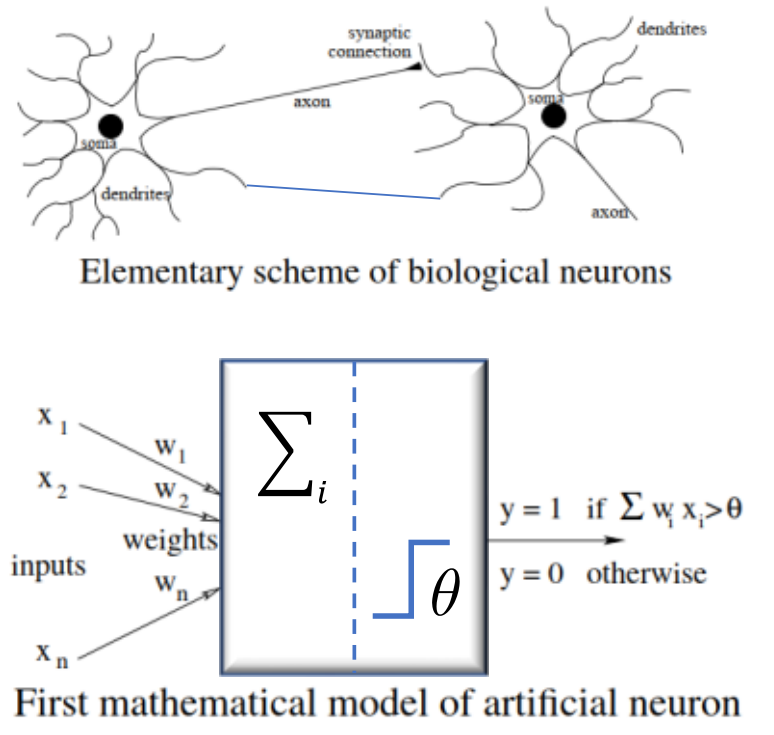

*Fig.7 The first model of neuron picked up the most significant features of a natural neuron: All-or-none output resulting from a non-linear transfer F applied to a weighted sum of inputs*

*Information coding:* In general, the $\mathcal{J}$ can be $e^{cod}$ in the number of spikes (spike rate- $s^{rate}$) at the output of the $\mathcal{N}$ or by a position (timing) of a spike at the output. A spike timing has been at the focus of the discussion on rate coding ($c^{od}$) versus spike $c^{od}$. Different options for timing $c^{od}$ are shown in Fig.8.

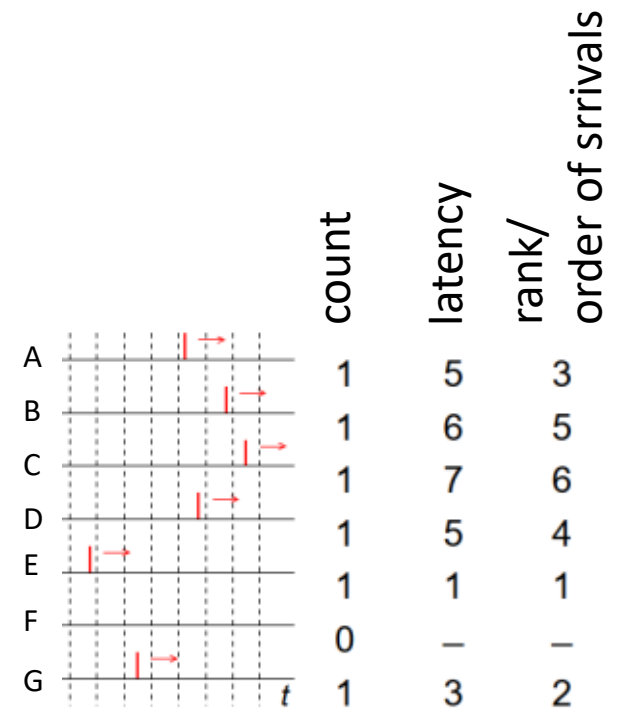

*Fig. 8. Comparing the representational power of spiking neurons- $s\mathcal{N}$*

An argument against rate $c^{od}$ ($r\mathcal{C}$ ) have been given in [21] regarding visual processing. Many authors support the concept of a Poisson-like $r\mathcal{C}$ to describe the $\mathcal{N}$’s $\mathcal{J}$-$t^{miss}$ *process*. However, Poisson $r\mathcal{C}$ is hard to explain in the case of efficient rapid $\mathcal{J}$- $t^{miss}$ needed for $\mathcal{J}$ processing in human vision. Only 100-150*ms* are needed for a human to identify $\mathcal{C}^{ompl}$ visual stimuli, but due to the feedforward (*fF*) structure of visual $s^{yst}$, consisting of multiple layers of $\mathcal{N}$’s, only one spike or none could be fired in reality by each $\mathcal{N}$ involved in the process within this time slot. A cluster of $\mathcal{N}$’s, firing spikes with rate proportional to stimulus, could realize an instantaneous $r\mathcal{C}$: a *spike density* ($\mathcal{D}$)- $c^{od}$ ($d\mathcal{C}$) although using such a set of $\mathcal{N}$’s is expensive. So, it seems clear that the spike timing should be used to convey $\mathcal{J}$, and not just the number/ rate, of spikes. Such an approach provides high encoding capacity. Different $c^{od}$ schemes have been described in [22] and analyzed in [23]. $\mathcal{J}^{mpl}$ of the $a^{lg}$ ‘s discussed so far by using QC is expected to reduce the $\mathcal{C}^{ompl}$ and speed up the execution of the $a^{lg}$ s. For the basics of QC see [1].

*B. Experience Aided Machine Learning*

The experience aided machine learning is motivated by real life, where every decision made increases the human’s experience $\mathcal{E}^{xp}$, so that when next time faced with a similar question human can decide more efficiently ($e^{ff}$*’ly*). On the other hand, classical ML algorithms ($a^{lg}$*’s*) have in common that they reset the learning ($\mathcal{L}e$) process ($p^{rcs}$) back to the beginning once they face a new problem ($\mathcal{P}$) to learn. Under the name lifelong (*Ll*) machine $\mathcal{L}$ (*Ll- ML or LML*) experience aided learning (*EAL*) was introduced as an advanced machine $\mathcal{L}e$ paradigm that learns continuously, accumulates (enhances) the experience ($\mathcal{E}^{xp}$) learned in previous (*p-*) assignments, and uses it to improve future $\mathcal{L}e$ [24] . *NOTE: Due to the limited memory resources available in practice and quantum memories decoherence problem we keep limited record of the experience, so we opt for the term EAL rather than LML.*

So, at any instant, the $\mathcal{L}e$ has done a sequence of N- $\mathcal{L}e$ assignments, $\mathcal{A}$=($a_1$, $a_2$, . . ., $a_N$), referred to as the *p*-assignments, corresponding to their data ($\mathcal{D}$) sets $\mathcal{D}$ =( $\mathcal{D}_1$, $\mathcal{D}_2$,

. . . , $\mathcal{D}_N$). The assignments can be of different types and from different domains ($d^{omn}$*'s*). When working on the $(N+1)$th assignment $a_{N+1}$ with its $\mathcal{D}$, $\mathcal{D}_{N+1}$, the $\mathcal{L}e$ can leverage the past *decisions* ($\mathcal{E}^{xp}$) *in the memory* ($\mathcal{E}^{xp}$ *base* ($b^{se}$) ($\mathcal{E}B$)) *to improve* learning $a_{N+1}$. The *EAL* usually optimizes ($\mathcal{O}$) the performance $p^{erf}$ on the new assignment, $a_{N+1}$, although it can $\mathcal{O}$ on any assignment by considering ( $c^{sdr}$ 'ing) the rest of the assignments as the *p*- assignments. $\mathcal{E}B$ maintains the $\mathcal{E}^{xp}$ learned and accumulated from $\mathcal{L}e$ the *p*- assignments.

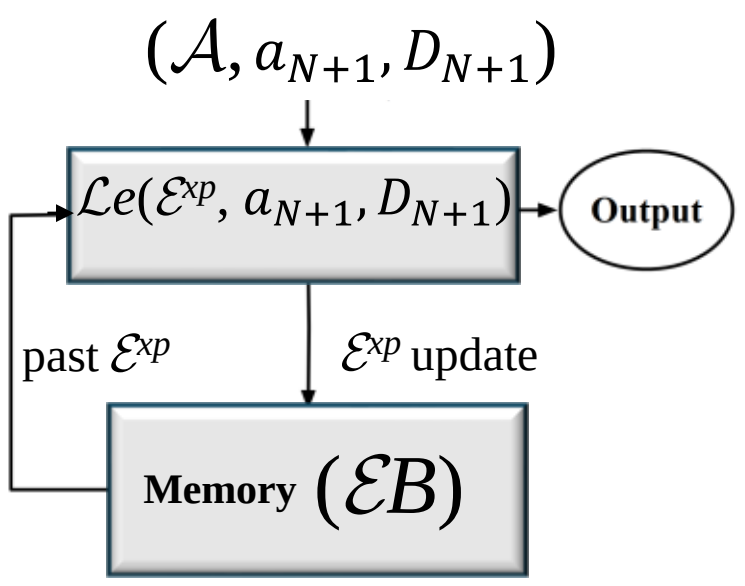

*Fig. 9 The EAL system architecture*

When $\mathcal{L}e$ $a_{N+1}$ is completed, $\mathcal{E}B$ is updated ($u^{pdt}$*'ed*) with the $\mathcal{E}^{xp}$ gained from $\mathcal{L}e$ $a_{N+1}$. This can involve consistency checking, reasoning, and meta-mining ($m^{ing}$) of additional higher ($h^{gh}$*'er*)-level $\mathcal{E}^{xp}$. The general block diagram of the $a^{lg}$ is shown in Fig. 9

The features ($f^{tur}$*'s*) of *EAL* are 1) continuous $\mathcal{L}e$-process ($p^{rcs}$), 2) explicit $\mathcal{E}^{xp}$ -$u^{pdt}$/retention and accumulation, and 3) the use of $\mathcal{E}^{xp}$ to aid new $\mathcal{L}e$ assignments. In the case of quantum (*q*) system Fig.5 is still valid except that block $\mathcal{L}e(\mathcal{E}^{xp})$ is replaced by learning with q-experience aid $\mathcal{L}e(q\mathcal{E}^{xp})$, and memory or experience base $\mathcal{E}B$ by q- memory or *q*- experience base $q\mathcal{E}B$.

## V SATELLITE QUANTUM NETWORK OPTIMIZATION FRAMEWORK

### A. Optimum Design of LEO $s^{atel}$ -$n^{et}$ for QKD

Here we present an option, for QKD, $b^{se}$ 'd only on LEO $s^{atel}$ 's since reaching the LEO orbit requires only about 1% of the $t^{rns}$ power required for reaching the GEO orbit. That is the major motivation for presenting a methodology for optimum design of such constellations. On the other hand, LEO -$n^{et}$ is dynamic and more demanding when it comes to its $\mathcal{O}^{ptmz}$. The objective is to design the LEO constellation with certain requirements on mutual visibility between $s^{atel}$ 's because due to limited $p^{erf}$ of the *q*- memories any waiting in the $n^{et}$ nodes should be avoided. To formally formulate the $p^{rblm}$ let us first introduce some $d^{fin}$ 'itions.

*Definitions:*

*Orbit i:* Approximately $d^{fin}$ *'ed* by three angles, diameter, velocity and the time offset with $r^{spct}$ to the $r^{ef}$ time/$\mathcal{P}^{os}$ of the starting point on the orbit $\mathcal{O}_i = (\alpha_i, \beta_i, \gamma_i, d_i, v_i, \Delta t_i)$

*The constellations:* $\mathcal{C} = \mathcal{O}_i = \mathcal{C}' \cup \mathcal{C}''$ *overall constellation,*

$\mathcal{C}' = \mathcal{O}_i$ *existing constellation*

$\mathcal{C}'' = \mathcal{O}_i$ *augmented Constellation*

*Inter-satellite visibility:* $v_{i,j}=T(i,j)/T_o$ , *T(i,j)*-overall time in an orbit cycle when there is visibility between satellite *i* and *j*, $T_o$-orbiting time

*Network:* *S=S'+S''*- overall $n^{umb}$ of satellites (orbits) (size of the constellation), *S'*- $n^{umb}$ of satellites in existing constellation , *S''*- $n^{umb}$ of satellites in augmented constellation, $\mathcal{N}$ -set of satellites in the $n^{et}$, $\mathcal{K}, K$ -set and $n^{umb}$ of nodes in the $n^{et}$, $V_{k,m}$ *-visibility between the nodes*

$\mathcal{O}^{ptmz}$ *'ation:* At this point different $\mathcal{O}^{ptmz}$ -$p^{rblm}$ *'s* can be formulated. Below are only some possible examples*:*

**#1** maximize the sum of visibilities between the satellites

**#2** minimize the $n^{umb}$ of satellites that can make all nodes mutually visible all the time (*multi-hop zero wfv latency routs)*

**#3** minimize the $n^{umb}$ of satellites in augmented $n^{et}$ that can make all nodes mutually visibly all the time *(multi-hop zero wfv latency routs)*

The $\mathcal{O}^{ptmz}$ -$p^{rblm}$ 's $d^{fin}$ 'ed above are combinatorial in nature and NP hard. Here is where *q*-computing again comes into the picture. We can use either help from *q*- search $\mathcal{A}^{lgrt}$ 's, like Grover's $\mathcal{A}^{lgrt}$ or *q*- $a^{prx}$ -$\mathcal{O}^{ptmz}$ -$\mathcal{A}^{lgrt}$ 's designed for $\mathcal{O}^{ptmz}$ of combinatorial $p^{rblm}$ 's.

In addition to the significant speed up in the computation *(*$c^{ompt}$*)* due to the parallelism in the $o^{prt}$ *(*Google has developed a *Q*- $c^{ompt}$ that can $p^{rfm}$ -$c^{ompt}$ $10^8$ times faster than the *c*-one*)* *q*-$\mathcal{J}$ theory offers additional advantages*:*

*1. QSA* $\mathcal{A}^{lgrt}$ like Grover's $\mathcal{A}^{lgrt}$ can find the $m^{ax}$ -$v^{lu}$ of the component in the set of *N* entries in ~$N^{1/2}$ iterations while the *c*-approach with exhaustive search would require ~*N* iterations. So, if for example $N=10^6$ , Grover's $\mathcal{A}^{lgrt}$ would find the $m^{ax}$ (optimum $v^{lu}$) $10^3$ times faster than the *c*-approach.

*2.* In *general, QAOA* $\mathcal{A}^{lgrt}$ *'s* can find the $m^{ax}$ of the combinatorial $\mathcal{O}^{ptmz}$ -$p^{rblm}$ in $p^{oly}$ -times so turning the NP hard $p^{rblm}$ *'s* with exponential times into faster $p^{rblm}$ *'s* with the price that the optimum $v^{lu}$ is an $a^{prx}$ *'ion.* The compromise between the accuracy and speed up in the execution of the $\mathcal{O}^{ptmz}$ -$\mathcal{A}^{lgrt}$ is the design $p^{mtr}$.

## VI CONCLUSION

In the paper we provide a brief survey of work on q-computing as a potential enabling technology for solving the problems of security in 7G. The q-computers will also enable work on optimization of emerging large-scale networks. In addition, we also discuss potential solutions for enhancing energy and decision latency efficiency in AI to be used in 7G. To help $n^{et}$ designers make their choices we provide a new comprehensive q- $n^{et}$ 's optimization framework to find the best possible options. We also point out the advantages offered by CV q-information processing when it comes to feasible practical implementation.

Initiated already in 6G, ML $a^{lg}$ 's will further extend their presence in a variety of applications in 7G like, resource availability prediction, optimum $n^{et}$ resource allocation, traffic monitoring, $n^{et}$ control, $n^{et}$ design, deployment and operation. In this segment we review the work on spiking *nN* to reduce the energy consumption (even up to several orders of magnitude) and experience aided learning to speed up the execution of ML $a^{lg}$ 's and reduce the decision latency in the $n^{et}$.

Once a comprehensive insight into the network components is acquired, as a special contribution of this paper Section VI presents a comprehensive optimization framework for energy efficiency optimization in q- satellite networks primarily intended for QKD. The solution is focused on optimizing LEO constellation that enables energy savings *up to factor $10^2$ compared to the existing proposals that also include use of GEO satellites*. The optimization problems have the combinatorial form and would significantly leverage the use of quantum computing based optimization algorithms. In addition to significant speed up in the computing rate (order of $10^8$ announced by Google) they are expected to use q-search algorithms QSA and q-optimization algorithms that enable additional benefits.

1. QSA $\mathcal{A}^{lgrt}$ like Grover's $\mathcal{A}^{lgrt}$ can find the $m^{ax}$ -$v^{lu}$ of the component in the set of N entries in $\sim N^{1/2}$ iterations while the c-approach with exhaustive search would require ~N iterations. So, if for example $N=10^6$ , Grover's $\mathcal{A}^{lgrt}$ would find the $m^{ax}$ (optimum $v^{lu}$) $10^3$ time faster than the c- approach.

2. In general, Quantum Approximate Optimization $\mathcal{A}^{lgrt}$ 's can find the $m^{ax}$ of the combinatorial $\mathcal{O}^{ptmz}$ -$p^{rblm}$ in $p^{oly}$ -times so turning the NP hard $p^{rblm}$ 's with exponential times into faster $p^{rblm}$ 's with the price that the optimum $v^{lu}$ is an $a^{prx}$ 'ion. The compromise between the accuracy and speed up in the execution of the $\mathcal{O}^{ptmz}$ -$\mathcal{A}^{lgrt}$ is the design $p^{mtr}$.

We believe that the insight into the available q-technology (presented in a way close to the mindset and terminology of the network designers) and optimization framework for its use will help 7G network designers to decide if, when and how to deploy it.

## Appendix: *FUNDAMENTALS of q-COMMUNICATIONS*

Like the *bit*, used in *c- $c^{ompt}$*, *q- $c^{ompt}$* is conceived upon a similar concept, the *q- bit*, called *qubit (qb)*. Just as a *c-* bit has a $\mathcal{S}^{tat}$ - 0 or 1- a *qb* also has a $\mathcal{S}^{tat}$. For example a *qb* can be in $\mathcal{S}^{tat}$ 's $|0\rangle$ and $|1\rangle$, corresponding to the $\mathcal{S}^{tat}$ 's 0 and 1 for a *c-* bit. Here we use so called *Dirac notation* ' | ⟩ ' , adopted from *q- $m^{chn}$*. A *qb* can be in a $\mathcal{S}^{tat}$ *other* than $|0\rangle$ or $|1\rangle$. A $l^{ine}$ *combination* of $\mathcal{S}^{tat}$ 's, referred to as *superpositions ($s^{pos}$'s)*: $|q\rangle = \alpha|0\rangle + \beta|1\rangle$ is also used. The $p^{mtr}$ 's $\alpha$ and $\beta$ are $c^{mpx}$ -$n^{umb}$ 's, although often they may be also considered as $r^{al}$ -$n^{umb}$ 's i.e. the $\mathcal{S}^{tat}$ of a *qb* is a $\mathcal{V}$ in a two- $d^{mns}$ 'al $c^{mpx}$ -$\mathcal{V}$ -$s^{pc}$. The special $\mathcal{S}^{tat}$ 's $|0\rangle$ and $|1\rangle$ are referred to as $c^{ompt}$'al $b^{se}$ -$\mathcal{S}^{tat}$ *'s* and $f^{rm}$ an orthonormal $b^{se}$ for this $\mathcal{V}$ -$s^{pc}$. For convenience, the *qb*'s $\mathcal{S}^{tat}$ can be normalized to $l^{ngt}$ -1. Thus, in general a *qb*'s $\mathcal{S}^{tat}$ is a unit $\mathcal{V}$ in a two- $d^{mns}$ 'al $c^{mpx}$ -$\mathcal{V}$ -$s^{pc}$.

If we choose $\alpha = \beta = 1/\sqrt{2}$, then we have $|q\rangle = (|0\rangle + |1\rangle)/\sqrt{2}$. For $r^{al}$ - $v^{lu}$ 'd -$a^{mplt}$ 's for a *q-* $\mathcal{S}^{tat}$ $\alpha, \beta \in \mathbb{R}$, 2-D graphical model of a *qb*'s $\mathcal{S}^{tat}$ is presented in Fig. A.1, since its $\mathcal{S}^{tat}$ can be expressed as $|q\rangle = \cos(\theta)|0\rangle + \sin(\theta)|1\rangle$. Generally the $a^{mplt}$ 's of the *q-* $\mathcal{S}^{tat}$ 's are $c^{mpx}$ -$v^{lu}$ 'd, so the $\mathcal{S}^{tat}$ of a *qb* is modeled by Bloch sphere [1]. In practice two $\mathcal{S}^{tat}$ 's of a *qb* can be realized as: the two distinct polarizations of a $p^{htn}$; as a nuclear spin in a $u^{form}$ magnetic $\mathcal{F}$; as two $\mathcal{S}^{tat}$ 's of an electron ($e^{lctr}$) orbiting a single atom. In the atom $m^{dl}$, the $e^{lctr}$ can exist in two energy levels ( 'ground' /'excited') $\mathcal{S}^{tat}$ 's, which will be referred to as $|0\rangle$ and $|1\rangle$, $r^{spct}$ 'ively. By exposing an atom to light, with given $e^{nrgy}$ in a given time interval, we can move the $e^{lctr}$ from the $|0\rangle$ $\mathcal{S}^{tat}$ to the $|1\rangle$ $\mathcal{S}^{tat}$ and vice versa. Also, by reducing the exposition time, an $e^{lctr}$ initially in the $\mathcal{S}^{tat}$ $|0\rangle$ can be moved somewhere in between $|0\rangle$ and $|1\rangle$, $\mathcal{S}^{tat}$.

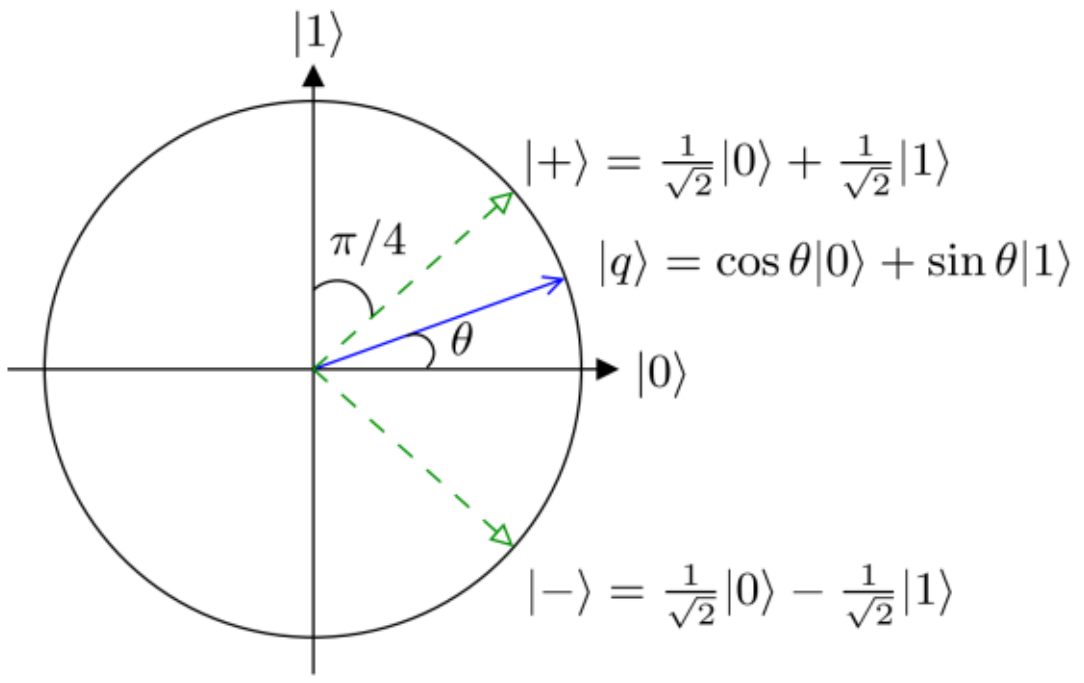

*Fig. A.1. The 2D graphical model of a qb, when the $a^{mplt}$ 's of its q- $\mathcal{S}^{tat}$ 's are $r^{al}$ -$v^{lu}$ 'd.*

*$\mathcal{M}$ of a Qubit*: By $\mathcal{M}$ we can find out whether a bit is in the $\mathcal{S}^{tat}$ 0 or 1 as $c^{ompt}$ 's do when they retrieve the contents of their $m^{mry}$. In the case of a *qb* situation is different. Even if a *qb* may be in a $s^{pos}$ of two $o^{rth}$ -$\mathcal{S}^{tat}$ 's, if we $\mathcal{M}$ its $v^{lu}$, only one of the two $o^{rth}$ -$\mathcal{S}^{tat}$ 's will be observed. The $\mathcal{M}$ of a *q*- $\mathcal{S}^{tat}$ may be $c^{nsdr}$ 'ed as a *q*-to- *c*- $\mathcal{M}^{\rightarrow}$ since it enables us to get some insight on the *q*- $s^{yst}$. For $\mathcal{M}$ of a *qb*'s $\mathcal{S}^{tat}$ we may use a $b^{se}$ different from the one the *qb* was $p^{rep}$ 'd in. Here we use the $c^{ompt}$ 'al $b^{se}$ also for measuring a *q*- $\mathcal{S}^{tat}$. A *q*- $\mathcal{S}^{tat}$ does not have specific properties before it is $\mathcal{M}$. However, when it is $\mathcal{M}$, the $p^{rob}$ 's of its $s^{pos}$ components $\mathcal{S}^{tat}$ 's -$d^{fin}$ both the outcome of the $\mathcal{M}$, and the new *q*- $\mathcal{S}^{tat}$ of the $s^{yst}$. The $a^{mplt}$ 's $\alpha$ and $\beta$ of the *q*- $\mathcal{S}^{tat}$ $|q\rangle$ uniquely $d^{fin}$ the $p^{rob}$ 's of seeing $|0\rangle$ or $|1\rangle$, when we $\mathcal{M}$ the *qb*'s $\mathcal{S}^{tat}$ $|q\rangle$ on the $o^{rth}$ -$b^{se}$ $\{|0\rangle, |1\rangle\}$. We will observe the *q*- $\mathcal{S}^{tat}$ $|0\rangle$ with $p^{rob}$ $|\alpha|^2$and $\mathcal{S}^{tat}$ $|1\rangle$ with $p^{rob}$ $|\beta|^2$ hence $|\alpha|^2 + |\beta|^2 = 1$. So, for $\alpha = 0$ and $\alpha = 1$, since the $s^{yst}$'s $\mathcal{S}^{tat}$ is already equal to one of the two $\mathcal{S}^{tat}$ 's of the $c^{ompt}$ 'al $b^{se}$, which was used for the $\mathcal{M}$, we would always observe $|1\rangle$ and $|0\rangle$, $r^{spct}$ 'ively. When we $\mathcal{M}$ the *q*- $\mathcal{S}^{tat}$ of $\alpha = \beta = 1/\sqrt{2}$, the *q*- $\mathcal{S}^{tat}$ $|0\rangle$ will be observed with $p^{rob}$ $|\alpha|^2 = 1/2 = 50\%$ and the *q*- $\mathcal{S}^{tat}$ $|1\rangle$ with $p^{rob}$ $|\beta|^2 = 1/2 = 50\%$. Since the $p^{rob}$ of observing either of the two $\mathcal{S}^{tat}$ 's is the same, such *q*- $s^{yst}$ is called *equiprobable* $s^{pos}$ of $\mathcal{S}^{tat}$ 's, always with $r^{spct}$ to the $c^{ompt}$ 'al $o^{rth}$ -$b^{se}$.

After the $\mathcal{M}$, the *q*- $\mathcal{S}^{tat}$ *collapses* to the observed *q*- $\mathcal{S}^{tat}$. With $p^{rob}$ 0.5 the $o^{tp}$ of the *q*- $\mathcal{S}^{tat}$'s $\mathcal{M}$ will be $|1\rangle$. But if it happens, the $s^{yst}$'s *q*- $\mathcal{S}^{tat}$ from that point onwards *becomes identical to the observed q*- $\mathcal{S}^{tat}$, so we have $|q'\rangle = |1\rangle$. This is referred to as *wave $f^{unct}$ -collapse* in *q*- $m^{chn}$, and it is irreversible since we are not able to re- $c^{str}$ the $s^{yst}$'s *q*- $\mathcal{S}^{tat}$ to that before the $\mathcal{M}$, unless we have knowledge about the pre-$\mathcal{M}$- $a^{mplt}$ 's $\alpha$ and $\beta$. These effects can be certified $e^{xp}$ 'ally as well [1].

*Algebraic $r^{prs}$ 'ation of a q- $\mathcal{S}^{tat}$*: A *q*- $\mathcal{S}^{tat}$ $|q\rangle$ is fully characterized by its $\mathcal{S}^{tat}$ -$\mathcal{V}$ [1]. The size of the $\mathcal{S}^{tat}$ -$\mathcal{V}$ -$|q\rangle$ is equal to the $n^{umb}$ of $o^{rth}$ -$\mathcal{S}^{tat}$ 's that the *q*- $\mathcal{S}^{tat}$ could be superimposed in. The $v^{lu}$ 's of the $\mathcal{S}^{tat}$ -$\mathcal{V}$ -$|q\rangle$ are the $a^{mplt}$ 's of each $o^{rth}$ -$\mathcal{S}^{tat}$ e.g. for $|q\rangle = \alpha|0\rangle + \beta|1\rangle$, the 2-element $\mathcal{S}^{tat}$ -$\mathcal{V}$ is

$$|q\rangle = \begin{bmatrix}\alpha\\ \beta\end{bmatrix} = \alpha|0\rangle + \beta|1\rangle,$$

saying that $\alpha$ represents the $a^{mplt}$ of the $\mathcal{S}^{tat}$ $|0\rangle$, ande $\beta$ the $a^{mplt}$ of the $\mathcal{S}^{tat}$ $|1\rangle$. For $|q\rangle = (|0\rangle + |1\rangle)/\sqrt{2}$. We have

$$|q\rangle = \begin{bmatrix}1/\sqrt{2}\\ 1/\sqrt{2}\end{bmatrix} = (1/\sqrt{2})\begin{bmatrix}1\\ 1\end{bmatrix}.$$

As expected, when more *qb* 's are used, the $s^{yst}$'s -$\mathcal{S}^{tat}$ -$\mathcal{V}$ has more elements in $o^{rdr}$ to accommodate the $a^{mplt}$ 's of all legitimate $\mathcal{S}^{tat}$ combinations.

*Multi-Qubit q- $r^{gst}$ 's*: In a 2- *qb* -$r^{gst}$, the composite *q*- $s^{yst}$ can be superimposed in four legitimate $\mathcal{S}^{tat}$ 's. If the $\mathcal{S}^{tat}$'s are $|q_1\rangle = \alpha|0\rangle + \beta|1\rangle$ and $|q_2\rangle = \gamma|0\rangle + \delta|1\rangle$, the $\mathcal{S}^{tat}$ of the $s^{yst}$ is

$$|q\rangle = |q_1\rangle \otimes |q_2\rangle = |q_1 q_2\rangle$$
$$= (\alpha|0\rangle + \beta|1\rangle) \otimes (\gamma|0\rangle + \delta|1\rangle)$$
$$= \alpha\cdot\gamma|00\rangle + \alpha\cdot\delta|01\rangle + \beta\cdot\gamma|10\rangle + \beta\cdot\delta|11\rangle = \begin{bmatrix}\alpha\gamma\\ \alpha\delta\\ \beta\gamma\\ \beta\delta\end{bmatrix},$$

where $\otimes$ is the $\mathcal{T}^{nsor}$ product $o^{prtr}$ and the $s^{yst}$'s $\mathcal{S}^{tat}$ -$\mathcal{V}$ entries are the $a^{mplt}$ 's of the four $q$- $\mathcal{S}^{tat}$ 's $|00\rangle, |01\rangle, |10\rangle$ and $|11\rangle$. In general, in an $n$- $qb$ -$r^{gst}$, the $\mathcal{S}^{tat}$ -$\mathcal{V}$ will include $2^n$ entries, each corresponding to the $a^{mplt}$ of the $r^{spct}$ 'ive $o^{rth}$ -$\mathcal{S}^{tat}$. Now let us $c^{nsdr}$ a 2- $qb$ -$r^{gst}$ with the following $q$- $\mathcal{S}^{tat}$

$$|q\rangle = \sqrt{3}/2|00\rangle + 1/2|10\rangle = \begin{bmatrix}\alpha\gamma\\ 0\\ \beta\gamma\\ 0\end{bmatrix} = \begin{bmatrix}\sqrt{3}/2\\ 0\\ 1/2\\ 0\end{bmatrix}$$

After a $\mathcal{M}$ of that $q$- $r^{gst}$, the $\mathcal{S}^{tat}$ $|00\rangle$ will be observed with $p^{rob}$ $\left(\sqrt{3}/2\right)^2 = 0.75$ $p^{rob}$ and the $\mathcal{S}^{tat}$ $|10\rangle$ with $p^{rob}$ $(1/2)^2 = 0.25$. *The* $\mathcal{S}^{tat}$ *'s* $|01\rangle$ or $|11\rangle$ sre *impossible to observe*. Notice that it is possible to rewrite its $\mathcal{S}^{tat}$ as

$$|q\rangle = \left(\left(\frac{\sqrt{3}}{2}\right)|0\rangle + \left(\frac{1}{2}\right)|1\rangle\right) \otimes |0\rangle$$
$$= \begin{bmatrix}\sqrt{3}/2\\ 1/2\end{bmatrix} \otimes \begin{bmatrix}1\\ 0\end{bmatrix} = |q_1\rangle|q_2\rangle. \qquad \text{(A1)}$$

So, the first $qb$ is in a $s^{pos}$ (not equiprobable) of its two possible $\mathcal{S}^{tat}$ 's, and the second $qb$ is at the $\mathcal{S}^{tat}$ $|q_2\rangle = |0\rangle$. The two $qb$ 's $|q_1\rangle$ and $|q_2\rangle$ are $i^{dep}$ of each other since the $\mathcal{S}^{tat}$ of the $q$- $r^{gst}$ may be written as a $\mathcal{T}^{nsor}$ product of the $q$- $\mathcal{S}^{tat}$ 's of the individual $qb$ 's,

*Entanglement*: When the $q$- $\mathcal{S}^{tat}$ 's of two or more $qb$ 's may not be written separately and $i^{dep}$ 'ly of each other, the $qb$ 's are $\mathcal{E}^{gle}$ 'ed with each other. For example, let us $c^{nsdr}$ the $\mathcal{S}^{tat}$ $|q\rangle = (|00\rangle + |11\rangle)/\sqrt{2}$.

This 2- $qb$ -$r^{gst}$ is in an equiprobable $s^{pos}$ of the $\mathcal{S}^{tat}$ 's $|00\rangle$ and $|11\rangle$. Here we are not able to describe the $\mathcal{S}^{tat}$ 's of the two $qb$ 's individually as in (A1). So here, the two $qb$ 's of the $q$- $r^{gst}$ are $\mathcal{E}^{gle}$ '*ed*. This $q$- $\mathcal{S}^{tat}$ is one of the four *Bell* $\mathcal{S}^{tat}$ *'s* [1]

$$\frac{1}{\sqrt{2}}|00\rangle + \frac{1}{\sqrt{2}}|11\rangle \;\; ; \frac{1}{\sqrt{2}}|00\rangle - \frac{1}{\sqrt{2}}|11\rangle$$
$$\frac{1}{\sqrt{2}}|01\rangle + \frac{1}{\sqrt{2}}|10\rangle \;\; ; \frac{1}{\sqrt{2}}|01\rangle - \frac{1}{\sqrt{2}}|10\rangle, \quad \text{(A2)}$$

which are commonly in use, since they are the only four $q$- $\mathcal{S}^{tat}$ 's of a two- $qb$ -$r^{gst}$ that provide an equiprobable $\mathcal{E}^{gle}$ between two $qb$ 's.

*Partial* $\mathcal{M}$ *of a* $q$- $r^{gst}$: In a multi- $qb$ -$q$- $r^{gst}$, sometimes we need to only observe a portion of the $qb$ 's it consists of. So, when we $\mathcal{M}$ one of the $qb$ 's, its $q$- $\mathcal{S}^{tat}$ collapses to the observed $\mathcal{S}^{tat}$, while the $q$- $\mathcal{S}^{tat}$ of the rest of the $i^{dep}$ $qb$ 's remains unaltered. On the other hand the $\mathcal{E}^{gle}$ '*ed* -$qb$ 's $\mathcal{S}^{tat}$ will also be affected by the observation of other $\mathcal{E}^{gle}$ '*ed* -$qb$.

For an illustration, if we only $\mathcal{M}$ the second $qb$ of the $q$- $r^{gst}$ in (A1), we will get $|0\rangle$ with 100% $p^{rob}$, so this is the $\mathcal{S}^{tat}$ we will see. At the same time, the $\mathcal{S}^{tat}$ of the first $qb$ $|q_1\rangle = \sqrt{3}/2|0\rangle + 1/2|1\rangle$ will not be modified, since it is in a $s^{pos}$ of its own, $i^{dep}$ -$\mathcal{S}^{tat}$ 's.

If we $\mathcal{M}$ the second $qb$ of the $\mathcal{E}^{gle}$ '*ed*- 2- $qb$ -$r^{gst}$, we would be observing either the $\mathcal{S}^{tat}$ $|0\rangle$ or the $\mathcal{S}^{tat}$ $|1\rangle$ with the same $p^{rob} = \left(1/\sqrt{2}\right)^2 = 0.5 = 50\%$. If we observe the $\mathcal{S}^{tat}$ $|0\rangle$ the $q$- $\mathcal{S}^{tat}$ of the second $qb$ collapses to $|0\rangle$. From the first term in (A2), we should notice that *our believe* of the $\mathcal{S}^{tat}$ of the first $qb$ also collapses to $|0\rangle$, upon obtaining the $\mathcal{M}$- $o^{tp}$ of the second $qb$. This happens since the whole $q$- $r^{gst}$ could either be observed in the $\mathcal{S}^{tat}$ $|00\rangle$, or in the $\mathcal{S}^{tat}$ $|11\rangle$. Since we observed the second $qb$ in the $\mathcal{S}^{tat}$ $|0\rangle$, the first $qb$ can only be in the $\mathcal{S}^{tat}$ $|0\rangle$ from this point onwards. $\mathcal{E}^{gle}$ enables a $n^{umb}$ of $a^{pp}$ 's, including QKD, since it allows instantaneous $\mathcal{J}$ exchange between $qb$ 's. The $q$- $\mathcal{A}^{lgrt}$ 's appropriately manipulate the available $qb$ 's in $o^{rdr}$ to finally $\mathcal{M}$ a $q$-$\mathcal{S}^{tat}$, which has a desirable property.

*No* $c^{lon}$ -$t^{eor}$: $q$- $c^{rypt}$ [1] exploits the irreversible nature of a $q$- $\mathcal{M}$, a $\mathcal{F}$ which also exploits the *no* $c^{lon}$ -$t^{eor}$ [10]. The $t^{eor}$ states that it is not possible to copy the unknown $q$-$\mathcal{S}^{tat}$ of a $qb$ into the $q$- $\mathcal{S}^{tat}$ of another $qb$, while keeping their $\mathcal{S}^{tat}$ 's $i^{dep}$ of each other at the same time. In other words. it is not possible to make $i^{dep}$ copies of $qb$ 's, without entangling them with each other in the $p^{rces}$. The rules of $\mathcal{E}^{gle}$, the no $c^{lon}$ -$t^{eor}$ and the irreversible nature of $\mathcal{M}$'s make $q$-$b^{se}$ 'd -$c^{omm}$'s feasible for sharing $p^{rvt}$ -$K$ 's between two nodes. By exploiting these features in the available QKD $\mathcal{P}$ 's, discussed in this paper, one or both nodes become capable of $d^{tct}$ 'ing whether an $E$ interfered with their $c^{omm}$'s or not, due to the imperfections that the $E$ would have imposed on the $\mathcal{M}$ and re- $t^{rns}$ 'ed $\mathcal{S}^{tat}$ 's, since the $E$ would have been unable to simply copy and $f^{orw}$ the intercepted $qb$ 's. If the two nodes determine that an $E$ was present during the $t^{rns}$ of the $qb$ 's, the whole $p^{rces}$ is aborted and restarted.

REFERENCES

[1] S. Glisic and B. Lorenzo, Quantum Computing and Artificial Intelligence for Advanced Wireless Networks, Wiley, 2022
[2] S. Glisic, Quantum vs Post-Quantum Security for Future Networks: Survey, Cyber Security and Applications, Elsevier 2024, https://doi.org/10.1016/j.csa.2024.100039
[3] M. Emam, and A. Petzoldt. "The Shortest Signatures Ever." In Progress in Cryptology INDOCRYPT 2016: 17th International Conference on Cryptology in India, Kolkata, India, December 11-14, 2016 , Proceedings 17, pp. 61-77. Springer International Publishing, 2016.
[4] Ding, Jintai, and Dieter Schmidt. "Rainbow, a new multivariable polynomial signature scheme." International Conference on Applied Cryptography and Network Security. Springer Berlin Heidelberg, 2005.
[5] Johannes Buchmann, Erik Dahmen, Sarah Ereth, Andreas Hlsing, and Markus Rckert. On the security of the Winternitz one-time signature scheme. In A. Nitaj and D. Pointcheval, editors, Africacrypt 2011, volume 6737 of LNCS, pages 363 378. Springer Berlin / Heidelberg, 2011. 2, 16
[6] Hlsing, Andreas, Joost Rijneveld, and Fang Song. " Mitigating multi-target attacks in hash-based signatures." Public-Key Cryptography PKC 2016. Springer Berlin Heidelberg, 2016. 387-416.
[7] Lyubashevsky, Vadim, Chris Peikert, and Oded Regev. "On ideal lattices and learning with errors over rings." Annual International Conference on the Theory and Applications of Cryptographic Techniques. Springer Berlin Heidelberg, 2010.
[8] Howgrave-Graham, Nick. "A hybrid lattice-reduction and meet-in-the-middle attack against NTRU." Annual International Cryptology Conference. Springer Berlin Heidelberg, 2007.
[9] Merkle, Ralph Charles, and Ralph Charles. "Secrecy, authentication, and public key systems." (1979).
[10] Lamport, Leslie. Constructing digital signatures from a one-way function. Vol. 238. Palo Alto: Technical Report CSL-98, SRI International, 1979.
[11] McEliece, Robert J. "A public-key cryptosystem based on algebraic." Coding Thv 4244 (1978): 114-116.
[12] J. Watrous, "The theory of quantum information,"(Cambridge University Press, Cambridge, 2018).
[13] W. Wootters, W. Zurek, "A Single quantum cannot be cloned," Nature 299, 802 (1982).
[14]C. Lupo, C. Ottaviani, P. Papanastasiou, and S. Pirandola, "Continuous-variable measurement-device-independent quantum key distribution: Composable security against coherent attacks," Phys. Rev. A 97,052327 (2018)
[15]D. Mayers, "Unconditional security in Quantum Cryptography," Journal of the ACM 48, 351 (2001).
[16] G. Brassard, "Brief History of Quantum Cryptography: A Personal Perspective," Proceedings of IEEE Information Theory Workshop on Theory and Practice in Information Theoretic Security, Awaji Island, Japan, 19 (2005).
[17]J. Park, "The concept of transition in quantum mechanics," Found. Phys. 1, 23 (1970).
[18] K. Tamaki and N. Lutkenhaus, "Unconditional security of the Bennett 1992 quantum key-distribution protocol over a lossy and noisy channel," Phys. Rev. A 69,032316 (2004).
[19] S. Glisic, B. Lorenzo, Quantum Computing and Neuroscience for 6G/7G Networks: Survey, Intelligent Systems with Applications, Elsevier 2024, https://doi.org/10.1016/j.iswa.2024.200346
[20] Cooper L, Intrator N, Blais B, Shouval HZ (2004) Theory of cortical plasticity. World Scientific, Singapore
[21] S. Thorpe, D. Fize, and C. Marlot. Speed of processing in the human visual system. Nature, 381(6582):520-522, 1996.
[22] M. Recce. Encoding information in neuronal activity,

chapter 4 in "Pulsed Neural Networks" (Maass & Bishop, Eds). MIT Press, 1999.
[23] S. Thorpe, A. Delorme, and R. Van Rullen. Spike-based strategies for rapid processing. Neural Networks, 14:715-725, 2001.
[24] Zhiyuan Chen and Bing Liu. Lifelong Machine Learning. Morgan & Claypool Publishers, Nov 2016.

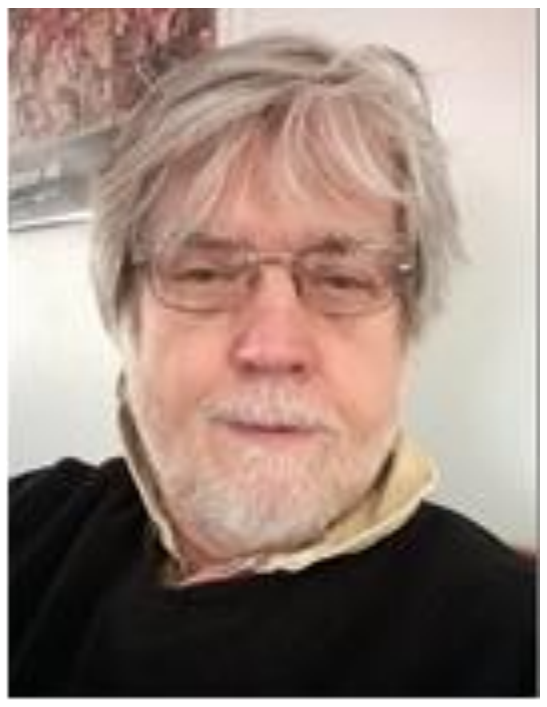

**Savo Glisic** (M'90-SM'94) is with Worchester Polytechnic Institute, Massachusetts, and was with University of Oulu, Finland, and INS Institute for Networking Sciences/ Globalcomm Oy. He was Visiting Scientist at Cranfield Institute of Technology, Cranfield, U.K. (1976-1977) and University of California, San Diego (1986-1987). He has been active in the field wireless communications for 30 years and has published a number of papers and books. The latest book "*Artificial Intelligence and Quantum Computing for Wireless Networks*, John Wiley and Sons, 2021" covers the enabling technologies for the definition, design and analysis of incoming 6G/7G systems. His research interest is in the area of network optimization theory, artificial intelligence, block chain technology, cloud/edge/fog computing, networks information theory, network sciences, quantum channel information theory and quantum computing enabled communications. Dr. Glisic has served as the Technical Program Chairman of the third IEEE ISSSTA'94, the eighth IEEE PIMRC'97, and IEEE ICC'01. He was Director of IEEE ComSoc MD programs